# Violation of Kohler's rule in $Ta_2PdTe_6$ and absence of same in $Nb_2PdS_5$: A high field magneto transport study

Reena Goyal[1,2], Rajveer Jha[2] and V.P.S. Awana[2]
reenagoyal88@gmail.com

[1]AcSIR-Academy of Scientific & Innovative Research- National Physical Laboratory, New Delhi-110012

[2]Quantum Phenomena and Applications, National Physical Laboratory (CSIR), New Delhi-110012, India

**Abstract.** Here, we present the comparative study of magnetotransport properties of recently discovered $Ta_2PdTe_6$ and $Nb_2PdS_5$ superconductors. The XRD and magnetotransport measurements are performed on these samples to investigate structure and superconducting properties as well as normal state transport properties of these compounds. Both the compounds are crystallized in monoclinic structure within space group C 2/m. Here, we observe superconductivity in both the compounds $Ta_2PdTe_6$ ($T_c$ =4.4 K) and $Nb_2PdS_5$ ($T_c$ =6.6 K). We see a linear magnetoresistance in $Ta_2PdTe_6$ as well as violation of Kohler's rule in same compound. On the other hand, we find the absence of same in $Nb_2PdS_5$ compound.

## INTRODUCTION

The recently discovered Quasi 1 D chain containing compounds $(Nb/Ta)_2Pd_x(S/Se/Te)_y$ have attracted lot of attention in scientific community due to their low dimensional structures. $Ta_4Pd_3Te_{16}$ compound has been reported to be superconducting with $T_c$=4.5 K.[5] The several parameters such as structure, upper critical field and charge carrier type are important for understanding superconducting mechanism in any new superconductor. It has been shown in earlier reports that the significant impact of spin orbit coupling along with multiband effect on large upper critical field [3]. The first principle calculation of electronic structure provides evidence of multiband nature of $Nb_2PdS_5$ superconductor [1]. The normal state property i.e. magnetoresistance is very important in understanding the behavior of charge carriers [7]. Recently, Xiao feng et. al. have showed the presence of quasi linear magnetoresistance and violation of Kohler's rule in $Ta_4Pd_3Te_{16}$ compound.[6]

In this context, we have compared the magnetoresistance in normal state of $Ta_2PdTe_6$ and $Nb_2PdS_5$ compound. We have observed the presence of quasi linear magnetoresistance in $Ta_2PdTe_6$ while absence of same in $Nb_2PdS_5$ compound. Additionally, we have also found the violation of Kohler's rule in $Ta_2PdTe_6$ compound.

## EXPERIMENTAL

Polycrystalline $Ta_2PdTe_6$ and $Nb_2PdS_5$ compounds have been synthesized via solid state reaction route followed by same procedure described in earlier reports [2, 4]. Room temperature XRD patterns have been recorded with Rigaku X-ray diffractometer using Cu Kα line of 1.54184 Å. The electrical measurements and detail study of magneto resistance in normal state have been performed with Quantum design Physical property measurement system (PPMS) equipped with 14 T superconducting magnet.

## RESULTS AND DISCUSSION

Figure 1 represents the room temperature X-Ray diffraction (XRD) patterns for synthesized $Ta_2PdTe_6$ and $Nb_2PdS_5$ compounds. The crystal structure of studied samples has been determined by Rietveld refinement of room temperature XRD data using Fullprof software. Both the compounds are found to be well crystallized in monoclinic C2/m structure. The XRD peaks are found to be well matched with our previous work on same compounds.

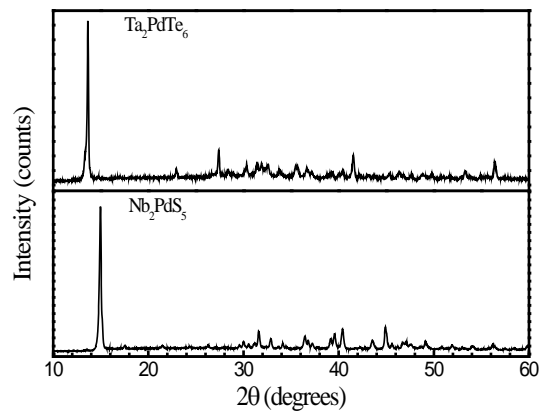

Figure 1: Room temperature XRD pattern for $Ta_2PdS_5$ and $Nb_2PdS_5$ compounds.

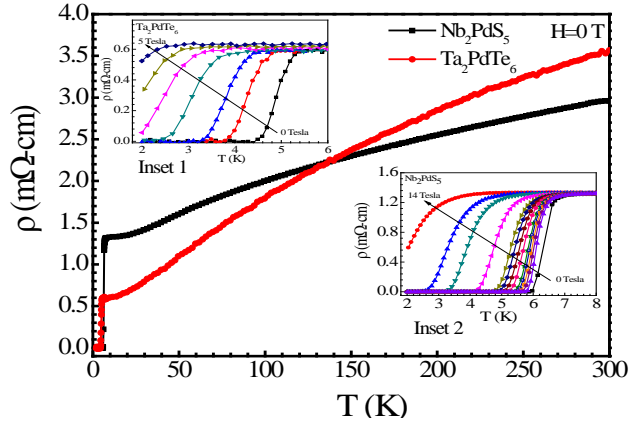

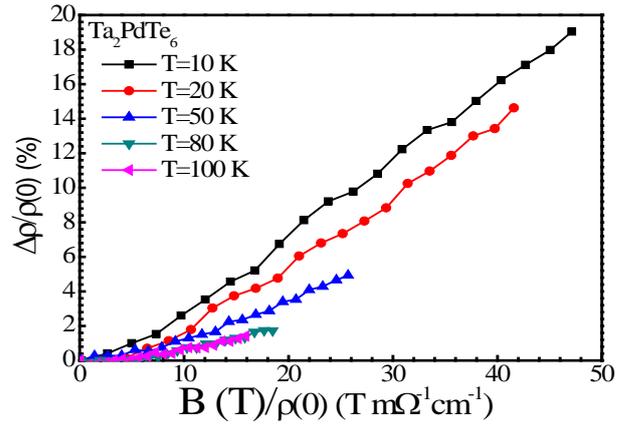

**Figure 2:** T dependence of resistivity ρ(T) for Ta$_2$PdTe$_6$ and Nb$_2$PdS$_5$ under zero applied magnetic field in range 2 K to 300 K. The inset 1 shows the same under different applied magnetic fields for Ta$_2$PdTe$_6$. The inset 2 shows the same under the applied magnetic field for Nb$_2$PdS$_5$.

Figure 2 represents the temperature dependence of resistivity for Ta$_2$PdTe$_6$ and Nb$_2$PdS$_5$ compounds. Ta$_2$PdTe$_6$ and Nb$_2$PdS$_5$ clearly exhibit a metallic behaviour with T$_c$ =4.4 K and T$_c$ =6.6 K where ρ=0. For both the compounds the zero resistivity point shifts clearly towards lower temperature with increase of applied magnetic field as shown in insets 1 and 2. Clearly, the effect of magnetic field is weak on Nb$_2$PdS$_5$ in comparison to Ta$_2$PdTe$_6$ compound.

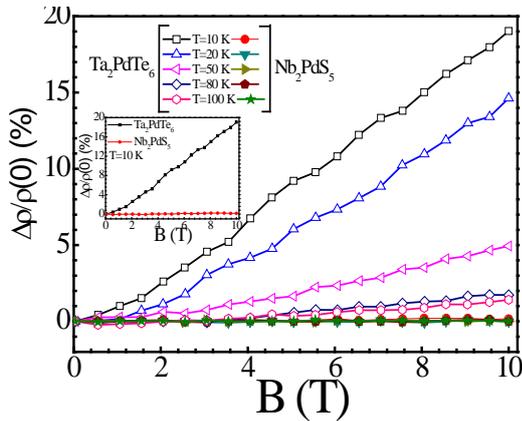

**Figure 3:** Field dependence of magnetoresistance for Ta$_2$PdTe$_6$ and Nb$_2$PdS$_5$ compounds at several temperatures. Inset shows same at T=10 K.

Figure 3 depicts the behaviour of ratio of magnetoresistance i.e. at different temperatures in normal state of Ta$_2$PdTe$_6$ and Nb$_2$PdS$_5$ compounds. Clearly, Ta$_2$PdTe$_6$ shows linear positive magnetoresistance at different temperatures and on the other hand Nb$_2$PdS$_5$ shows no change in resistance with applied field. Apparently, in Ta$_2$PdTe$_6$, the value of magnetoresistance increases with temperature at fixed value of magnetic field.

**Figure 4:** Kohler's plot for sample Ta$_2$PdTe$_6$ in a field range from 0 T to 10 T at several temperatures.

Kohler's plot for Ta$_2$PdTe$_6$ is illustrated in figure 4. Clearly, the data measured at different temperatures do not overlap and Kohler's rule is violated for Ta$_2$PdTe$_6$ compound. Since, Kohler's rule says that magnetoresistance, Δρ= [ρ (H) - ρ (0)]/ ρ (0), plotted against B/ρ (0) should give single curve for all temperatures. Kohler's rule was not observed in MgB$_2$ superconductor because of the multiband property.

CONCLUSIONS

We have successfully synthesized Ta2PdTe6 and Nb$_2$PdS$_5$ compound via solid state reaction route. The studied compounds Ta$_2$PdTe$_6$ and Nb$_2$PdS$_5$ show superconductivity at 4.4 K and 6.6 K. Also, Ta$_2$PdTe$_6$ compound shows linear magnetoresistance and absence of same in Nb$_2$PdS$_5$ compound. Also, the Kohler' rule is violated in Ta$_2$PdTe$_6$.